\documentclass[aps,showpacs,twocolumn,showkeys,amsmath,amssymb]{revtex4-1}

\usepackage[utf8]{inputenc}
\usepackage{amsmath,amssymb,amsfonts}
\usepackage{graphicx}
\usepackage{txfonts}
\usepackage{epstopdf}
\usepackage{eurosym}

\begin{document}

\title{An equilibrium-conserving taxation scheme for income from capital}

\author{Jacques Tempere}
\affiliation{Theory of Quantum and Complex Systems, Universiteit
  Antwerpen, Universiteitsplein 1, B-2610 Antwerpen, Belgium}
\date{\today}

\begin{abstract}
    Under conditions of market equilibrium, the distribution of capital
    income follows a Pareto power law, with an exponent that characterizes
    the given equilibrium. Here, a simple taxation scheme is proposed such 
    that the post-tax capital income distribution remains an equilibrium 
    distribution, albeit with a different exponent. This taxation scheme
    is shown to be progressive, and its parameters can be simply derived
    from (i) the total amount of tax that will be levied, (ii) the threshold
    selected above which capital income will be taxed and (iii) the total
    amount of capital income. The latter can be obtained either by using
    Piketty's estimates of the capital/labor income ratio or by fitting 
    the initial Pareto exponent. Both ways moreover provide a check on 
    the amount of declared income from capital.
\end{abstract}


\maketitle

\section{Introduction}

The distribution of income has been studied for a long time in the 
economic literature, and has more recently become a topic of investigation for statistical 
physicists turning to econophysics \cite{Chak1,Chak2,Yakovenko,Chak3}. 
The income distribution is characterized by a density function $f(x)$ such that $f(x)dx$ 
is the number of individuals earning income between $x$ and $x+dx$. 
From the empirical data obtained from tax records, two different regimes are 
readily distinguished. 
For income levels below a certain threshold $x_c$, the distribution 
follows and exponential (Boltzmann) law, $f(x) \propto \exp(-x/\bar{x})$, whereas for 
income levels above $x_c$ the distribution is better fitted by a power law, 
$f(x) \propto x^{-\gamma}$.
Note that for the bottom incomes, a deviation from the Boltzmann law is visible. 
This is due
to redistribution (such as social security benefits) which lifts a 
certain amount of people above a poverty threshold 
$x_{\textrm{pov}}$. 

For income above the poverty level but below $x_c$
the distribution is very well fitted by a Gibbs distribution 
\cite{Dragulescu2000,Ispolatov98}. Tax records that keep track of the 
source of income indicate that income in this regime is dominated by 
\textit{labor income} (salaries and wages) \cite{Willis}. In this regime, 
economic transactions can be modelled by additive processes \cite{Ispolatov98,Chak3}: 
money exchanges hands between agents but the total amount of money is conserved over the 
transaction. For example, each month an employee gets a certain sum of money added
to his account, and this sum is subtracted from the account of the employer's company.
Using this principle of local money conservation, Dragulescu and Yakovenko 
\cite{Dragulescu2000} have shown that the equilibrium distribution of
money over the agents involved in additive transactions follows a Boltzmann-Gibbs
exponential distribution. 
Note that this is a strongly simplified model of economic activity: 
it is clear that in reality global money conservation is violated. Indeed, banks can 
issue (or recall) loans, thereby increasing (or decreasing) the total money supply. 

This brings us to the higher incomes, $x>x_c$. As noted already by Pareto in 1897, 
these follow a power law \cite{Pareto}. Records that keep track of the source of 
income reveal that income in the power law regime is mainly \textit{capital income} 
(rent, profits, interests, dividends,...) \cite{Clementi,Hungerford}. 
For these types of income, economic transactions are better modeled by multiplicative
processes \cite{Ispolatov98,Levy2003,Fujiwara2003}. In contrast to a wage worker, a rentier 
expects that at each time step the money in his investment is multiplied by an interest 
factor. For such processes, it is $\log(x)$ rather than $x$ which is conserved locally, 
and this leads naturally to a power law rather than an exponential equilibrium 
distribution \cite{Levy1996}.


The change from exponential regime to power-law regime occurs in a narrow interval 
\cite{Dragulescu2001,Dragulescu2003} around $x_c$, allowing to separate not only 
the income in two sources ($M_{\textrm{lab}}$ from labor and $M_{\textrm{cap}}$ 
from capital), 
but also the population in two groups ($N_{\textrm{lab}}$ and 
$N_{\textrm{cap}}$, respectively). Of course, these are not clearly delineated, 
and also people filing tax forms for income below $x_c$ have a portion of their 
income coming from returnon capital, but the problem can be greatly simplified 
by taking the main source of income to be the entire income. The value $x_c$ 
separating the exponential from the power-law regimes lies between three and 
four times the average income in the exponential part of the 
distribution \cite{Dragulescu2001,Dragulescu2003,Silva2006}. 

The question that I wish to address in this paper is how to levy taxes such
that the immediate after-tax income distribution remains in equilibrium (i.e.
of Boltzmann type for $x<x_c$, and of Pareto type for $x>x_c$). The parameters
($\bar{x}$ and $\gamma$) of the pre-tax and after-tax distributions will be different, 
reflecting the change from one equilibrium to another rather than a shift to a 
non-equilibrium distribution. 
Free-market advocates can argue that this type of equilibrium-to-equilibrium
taxation is the least disruptive choice. The conjecture behind this is that when the 
income distribution is pushed strongly out of equilibrium the market is far from 
optimized as transactions that are wished for may not take place. Hence a taxation scheme 
that results in an out-of-equilibrium after-tax income distribution would be more 
detrimental to the market. Of course, whether any scheme is just or desirable is well beyond 
the scope of this paper. Nevertheless, given the current debate taking place 
(in the USA, the UK and the EU) of how to tax the rich, I believe the results 
presented here can contribute to an informed discussion.

\section{Capital and labor income for Belgium}

\begin{figure}
    \centering
    \includegraphics[width=8.6cm]{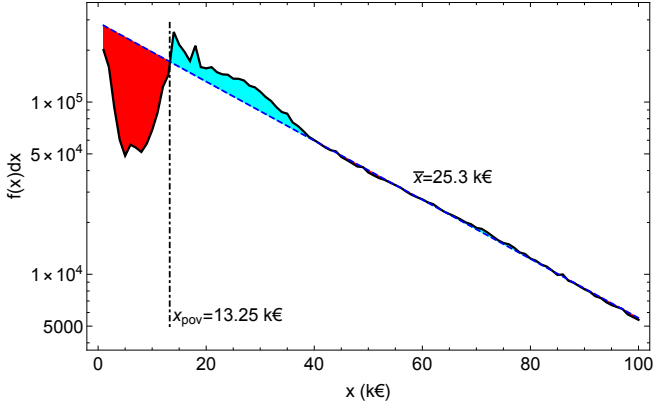}
    \caption{(color online) The (annual) income distribution in Belgium in 2014 as 
    obtained from personal tax records \cite{taxdata}, 
    in bins of d$x=1$ k\euro. The distribution follows a 
    Boltzmann law (dashed line) except for the lowest income. 
    Redistributive taxation lifts people with an income below $x_{\textrm{pov}}$ 
    (area shaded in dark red) above this poverty threshold (area shaded in light blue).}
    \label{fig1}
\end{figure}

To illustrate the results with numbers, I use my home country of Belgium as an
example, taking $x_c \approx 100$ k\euro. There were 
$N_{\textrm{tot}}=6.26 \times 10^6$ people filing a non-zero income 
tax record in Belgium in 2014, the latest year with complete information \cite{taxdata}.
Of these, 
$N_{\textrm{cap}}=1.73 \times 10^5$ (or 2.8\% of the total) indicate an income above $x_c$.
The remaining $N_{\textrm{lab}}=6.09 \times 10^6$ are in the exponential regime and 
represent a cumulative income of $M_{\textrm{lab}}=170.6$ G\euro. As seen in
Fig.~\ref{fig1}, there is a strong
deviation of the exponential regime for income levels below 
$x_{\textrm{pov}}=13.25$ k\euro : this is the poverty threshold. People below
this threshold receive social security benefits lifting them to roughly the
threshhold level and slightly above.

Estimating the amount of capital income $M_{\textrm{cap}}$ for Belgium is difficult 
to do based on tax forms. The reason is that not all sources of capital income need to be 
declared in Belgium: For instance, income from renting out apartments or offices 
is not declared. Piketty \cite{Piketty} provides an alternative way to estimate
$M_{\textrm{cap}}$ from $M_{\textrm{lab}}$. His study of the historical ratio
between capital income and labor income for developed economies shows that this
income has its own slow dynamics over time. Currently, the capital share of income 
in rich countries stands at 25-30\% of national income. The dynamics of the 
capital/labor income ratio is slow enough such that the income distribution is 
close to equilibrium at any time. Taking the above-mentioned capital share of 
the national income into account results in an estimate\cite{noot1} 
of $M_{\textrm{cap}} \approx 60$ G\euro.

The additive class, $x_{\textrm{pov}} < x < x_c$ is subject to the (normalized) Boltzmann-Gibbs 
distrubution
\begin{equation}
 f_{\textrm{lab}}(x)=\frac{N_{\textrm{lab}}}{\bar{x}} \exp(-x/\bar{x}),
\end{equation}
with $\bar{x}=M_{\textrm{lab}}/N_{\textrm{lab}}$ the average income in this class. 

The multiplicative class is subject to the power law distribution
\begin{equation}
 f_{\textrm{cap}}(x>x_c)= \frac{(\gamma-1) N_{\textrm{cap}}}{x_c} (x/x_c)^{-\gamma}. 
\end{equation}
The Pareto parameter $\gamma$ is fixed by $M_{\textrm{cap}}$ through the normalization
$M_{\textrm{cap}}=\int_{x_c}^{\infty}{x f_{\textrm{cap}}(x)dx}$ by
\begin{equation}
  \gamma=\frac{2 M_{\textrm{cap}} - N_{\textrm{cap}} 
  x_c}{M_{\textrm{cap}} - N_{\textrm{cap}} x_c}. \label{gam}
\end{equation}
This restricts $\gamma>2$, since $M_{\textrm{cap}}>N_{\textrm{cap}} x_c$. 
The estimate $M_{\textrm{cap}}=60$ G\euro ~ based on Piketty's observations 
corresponds to $\gamma=2.4$. This value is in agreement with the estimate
of $\gamma \approx 2.5$ obtained by Silva and Yakovenko
\cite{Silva2006}.  

Eq.~(\ref{gam}) can also be inverted, so that an empirical fit 
yielding $\gamma$ (in combination with $x_c$ and $N_{\textrm{cap}}$) may be 
used to estimate the total amount of income from capital. 
Detailed data for the high-income distribution is not publicly available in 
Belgium, and as mentioned above, exemption of some capital income sources
in Belgium complicates data finding. However, for countries with an obligation 
to declare all capital income, Eq.~\ref{gam} could be used to estimate the 
amount of undeclared income and hence the level of tax evasion by rentiers.
A similar proposal, based on deviations from the Pareto distribution, was 
introduced to estimate the size of shadow banking \cite{Marsili}.


\section{Taxing capital income}

Suppose one wants to levy taxes to raise a given amount $\Delta M$ of money. For
example, to bring all the poor to a minimum wage of $x_{\textrm{pov}}$, one would need
\begin{eqnarray}
  \Delta M &=& \int_{0}^{x_c} (x_c-x)f_{\textrm{lab}}(x)dx  \\
  &=& M_{\textrm{lab}} \left[ (x_{\textrm{pov}}/\bar{x})
  - (1-e^{-x_{\textrm{pov}}/\bar{x}}) \right].
\end{eqnarray}
Using the numbers listed above for Belgium, the topping up of all lower
incomes to $x_{\textrm{pov}}$ would require 16.4 G\euro.

Suppose moreover that one wants to obtain the amount of $\Delta M$ by taxing
capital income in such a way that the after-tax capital income distribution
follows again a Pareto law. The post-tax power law distribution necessarily 
has a different exponent $\eta > \gamma$, given by
\begin{equation}
  \eta = \frac{2 (M_{\textrm{cap}}-\Delta{M}) - N_{\textrm{cap}} 
  x_c}{M_{\textrm{cap}}-\Delta{M} - N_{\textrm{cap}} x_c}. 
  \label{eta}
\end{equation}
In our Belgian example, this would mean a change from $\gamma=2.41$ to
$\eta=2.66$. To describe the taxation scheme, we introduce a function
$X(x)$ that gives the post-tax net income $X$ as a function of the 
pre-tax income $x$. The distribution of post-tax income is denoted by
\begin{equation}
   f_{\textrm{cap}}^{\textrm{post-tax}}(X)= \frac{(\eta-1) 
   N_{\textrm{cap}}}{x_c} (X/x_c)^{-\eta}.
\end{equation}
This distribution has to obey
\begin{equation}
   f_{\textrm{cap}}^{\textrm{post-tax}}(X) dX = f_{\textrm{cap}}[x(X)] dx.
\end{equation}
Substituting the Pareto distributions in the above equation yields a 
differential equation for X(x),
\begin{equation}
   \frac{(\eta-1)}{x_c^{1-\eta}} X^{-\eta}(x) \frac{dX}{dx} =
   \frac{(\gamma-1)}{x_c^{1-\gamma}} x^{-\gamma}.
\end{equation}
It solution depends on the parameter 
\begin{equation}
  \tau = \frac{\gamma-1}{\eta-1}, \label{tau}
\end{equation}
and is given by
\begin{equation}
  X(x) = x_c^{1-\tau} x^{\tau}.
\end{equation}
As $\tau < 1$, this solution corresponds to a weighted geometric averaging 
between the capital income and the threshold $x_c$ where main income switches 
from additive to multiplicative. For a pre-tax income $x$ the 
corresponding tax rate $T(x)=(x-X)/x$ is given by
\begin{equation}
  T(x) = 1-\left( x_c / x \right)^{1-\tau}.  \label{taxrate}
\end{equation}
\begin{figure}
    \centering
    \includegraphics[width=8.6cm]{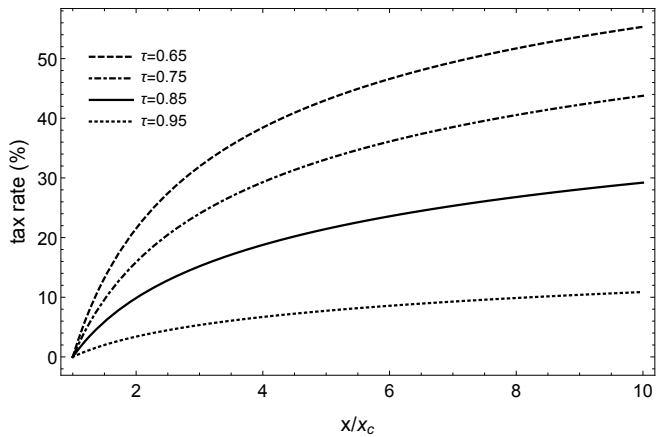}
    \caption{Taxation levels as a function of capital income, for different
    values of the parameter $\tau$, which is fixed by the amount of tax $\Delta M$
    that will be levied from the total capital income $M_{\textrm{cap}}$. The
    tax rate is progressive and taxation starts at the income level $x_c$ separating
    the Boltzmann-type labor income distribution from the Pareto-type capital
    income distribution.}
    \label{fig2}
\end{figure}
This represents a progressive tax rate. Taking our example
$\tau=0.85$, and the resulting tax rate is shown as the full curve in
Fig.~\ref{fig2}. Whereas for a capital income of 120 k\euro ~ (slightly
above the threshold when one can be called a rentier or ``rich'') 
the tax rate is about 3\%, at 200 k\euro ~ it has risen to 10\%, doubling
again to 20\% for 500 k\euro.

\section{Discussion and conclusions}
Eqs. (\ref{eta}),(\ref{tau}) and (\ref{taxrate}) represent a simple 
taxation scheme that preserves the power law nature of capital income. In
order to implement it, policy makers need to select a threshold $x_c$ of
income from capital above which the tax is levied, and choose a value of 
$\tau$, or equivalently, an amount $\Delta M$ of money that the tax should
raise. The basic idea behind
preserving the power law is that this law represents a distribution in
which the market is in equilibrium. 

Free market proponents claim, in the first fundamental theorem of
welfare economics, that a competetive market produces a (non-unique) Pareto 
efficient equilibrium outcome. Moreover, in this train of thought, a social 
planner could select the most suitable efficient outcome by lump sum transfers,
according to the second fundamental theorem. In this context, the current
proposal for taxation is precisely a way to organise a transfer that links one 
equilibrium for \textit{capital} income to another.

What if one would use the same logic to labor income? Changing one Boltzmann
distribution into another only requires a scale change $x \rightarrow \alpha x$
where $\alpha=1-\Delta M/M_{\textrm{lab}}$. This corresponds to a proportional
tax system (a ``flat tax'', such as a fixed sales tax). This is not always seen 
as the socially most desirable outcome as it penalizes the low-income segment 
of the population, who have less disposable income. In essence, the current
proposal represents a flat tax on $\log(x)$, modified by the 
presence of a the threshold $x_c$.
Regardless of the desirability debate, it is clear that proponents of a 
flat tax for labor income who base their
arguments on market efficiency, should then logically advocate the current 
progressive tax on capital income. Another commonly encountered argument for a 
flat tax is its simplicity. In this respect, the proposal for a progressive 
capital income taxation put forward in this paper offers a scheme which, at 
least to a physicist, is of similar simplicity.

\end{document}